\documentclass[3p,times,procedia]{elsarticle}
\usepackage{nupha_ecrc}
\volume{00}

\firstpage{1}

\journalname{Nuclear Physics A}

\runauth{}

\jid{nupha}

\jnltitlelogo{Nuclear Physics A}

\usepackage{amssymb}
\usepackage{lineno}

\usepackage[figuresright]{rotating}

\begin{document}
%\linenumbers
\begin{frontmatter}



%% Instructions from Editor: Please use the following \dochead only in the preprint version (e-print arXiv etc.); 
%% use empty \dochead{} when submitting to Nuclear Physics A!
\dochead{XXVIIth International Conference on Ultrarelativistic Nucleus-Nucleus Collisions\\ (Quark Matter 2018)}
%% e.g. \dochead{17th International Conference on Dynamical Processes in Excited States of Solids}

\title{Directed flow of quarks from the RHIC Beam Energy Scan measured by STAR}


\author{Gang Wang (for the STAR\fnref{col1} Collaboration)}
\fntext[col1] {A list of members of the STAR Collaboration and acknowledgements can be found at the end of this issue.}
\address{Department of Physics and Astronomy, University of California, Los Angeles, California 90095, USA}

\begin{abstract}
Directed flow ($v_1$) is a good probe of the early-stage dynamics of collision systems, and
the $v_1$ slope at midrapidity ($dv_1/dy|_{y=0}$)  is sensitive to the system's equation of state.
%Previously, a coalescence picture has been used to relate the elliptic flow ($v_2$) of identified particles to the $v_2$ of their constituent quarks. In this study, the coalescence idea is extended to $v_1$ of the constituent quarks versus $\sqrt{s_{NN}}$ --- this includes quarks produced in the collision, as well as those transported from the initial-state nuclei. 
STAR has published $v_1(y)$ measurements for ten particle species ($\pi^\pm$, $p$, $\bar{p}$,
$\Lambda$, $\bar{\Lambda}$, $\phi$, $K^\pm$ and $K^0_{S}$) in Au+Au collisions at eight beam
energies from $\sqrt{s_{NN}} = 7.7$ GeV to 200 GeV. In this study, we employ a simple coalescence idea to decompose $v_1$ of hadrons into $v_1$ of constituent quarks. The $dv_1/dy$ values of $\bar{p}$, $K^-$ and
$\bar{\Lambda}$ are used to test the coalescence sum rule for produced quarks.  
%We hypothesize that the $v_n$ of observed mesons and baryons is the summed $v_n$ of their constituent quarks, and it is assumed that deconfined quarks have already acquired azimuthal anisotropy.
Data involving produced quarks support the coalescence picture at $\sqrt{s_{NN}} = 11.5$
GeV to 200 GeV, and a sharp deviation from this picture is observed at 7.7 GeV.
The $dv_1/dy$ of transported quarks is studied via net particles (net $p$ and net $\Lambda$). In these proceedings, we further extract the $v_1$ slopes of produced and transported quarks, assuming that the coalescence sum rule is valid. 

\end{abstract}


\end{frontmatter}

%\linenumbers

\section{Introduction}
\label{intro}

Rapidity-odd directed flow ($v^{\rm odd}_1
(y)$) is the first Fourier coefficient of the final-state azimuthal distribution relative to the collision reaction plane~\cite{v1odd}, and describes the collective sideward motion of emitted particles. The rapidity-even component
($v^{\rm even}_1(y)$)~\cite{v1even} is unrelated to the reaction plane in symmetric
collisions, and hereafter, $v_1(y)$ implicitly
refers to the odd component. Hydrodynamic calculations~\cite{hydro1, hydro2} have proposed a minimum in net-baryon directed flow versus $\sqrt{s_{NN}}$ as a signature of a first-order phase transition between hadronic matter and quark-gluon plasma. This minimum is supposedly related to the softening of the system's equation of state (EOS).
STAR has published $v_1(y)$ measurements for ten particle species ($\pi^\pm$, $p$, $\bar{p}$,
$\Lambda$, $\bar{\Lambda}$, $\phi$, $K^\pm$ and $K^0_{S}$) in Au+Au collisions at eight beam
energies from $\sqrt{s_{NN}} = 7.7$ GeV to 200 GeV~\cite{STAR1,STAR2}. Net protons do show a minimum in $dv_1/dy$ near $\sqrt{s_{NN}}$ of 10 to 20 GeV~\cite{STAR1}.
In more recent model calculations of $v_1(y)$ with different EOS~\cite{th1,th2,th3,th4,th5,th6}, the assumption of purely hadronic physics is disfavored, but there is no
consensus on whether STAR measurements~\cite{STAR1} favor a
crossover or first-order phase transition. 
While further progress in models is needed for a definitive interpretation, the experimental data 
may distinguish different interpretations by advancing to the constituent quark level.

Number-of-constituent-quark (NCQ) scaling of elliptic flow ($v_2$)~\cite{v2} behaves as if $v_2$ is imposed at the level of deconfined constituent quarks, providing an example of coalescence behavior among quarks. The ten particle species available in the present analysis
allow a detailed investigation of the scaling behavior of $v_1$ at constituent quark level versus $\sqrt{s_{NN}}$.
We will test a set of assumptions, namely that $v_1$ is imposed at the pre-hadronic stage, that specific types
of quark have the same directed flow, and that the detected hadrons are formed via coalescence~\cite{v2,dunlop}. In a scenario where deconfined quarks have already acquired $v_n$, and in the limit of small  $v_n$, coalescence leads to the $v_n$ of the resulting mesons or baryons being the summed $v_n$ of their constituent quarks~\cite{dunlop,fries}. We call this assumption the coalescence sum rule. NCQ scaling in turn follows from the coalescence sum rule~\cite{dunlop}.
%, with the condition that $v_n$ is the same for the quarks involved. In the following test of the coalescence sum rule, we will invoke weaker conditions.

In this study, we attempt to separate ``transported'' quarks ($u$ and $d$ from the initial-state nuclei) ) and ``produced'' quarks ($u$, $\bar u$, $d$, $\bar d$, $s$ and $\bar s$ created in pair after the collision).
The number of transported quarks is conserved, and transported quarks 
%tagged with this quantum number 
experience the whole system evolution, including the initial sideward deflection, the possible softening of  equation of state, annihilation, re-scattering and so on. Conversely, the total number of produced quarks is not conserved, and produced quarks are presumably created in different stages~\cite{wave},
which complicates the interpretation of produced-quark $v_1$.
Experimentally, produced quarks can be studied with purely ``produced'' particles, such as $\bar p$, $\bar \Lambda$ and $K^-$, whereas transported quarks can be probed with net particles that represent
the excess yield of a particle species over its antiparticle. We define $v_{1\,{\rm net }\,p}$ based on expressing $v_1(y)$ for all protons as
$v_{1\,p} = r(y)v_{1\,{\bar p}} + [1-r(y)]v_{1\,{\rm net}\,p}$, where $r(y)$ is the ratio of observed $\bar p$ to $p$ yield at each beam energy. Net-$\Lambda$ $v_1$ is defined similarly, except $\bar p$ ($p$) becomes $\bar \Lambda$ ($\Lambda$).

\section{Results}
\label{results}
We first test the coalescence sum rule in a straightforward case where all quarks are known to be produced. Figure~1(a) compares the observed $dv_1/dy$ for $\bar \Lambda({\bar u}{\bar d}{\bar s})$ with the calculation for $K^-({\bar u}s) + \frac{1}{3}{\bar p}({\bar u}{\bar u}{\bar d})$~\cite{STAR2}. This calculation is based
on the coalescence sum rule combined with the assumption that $\bar u$ and $\bar d$ quarks have the same flow, and that $s$
and $\bar s$ have the same flow.
Close agreement is observed for $\sqrt{s_{NN}}$ from $11.5$ to $200$ GeV. The inset in Fig.~1(a) presents the same comparison, but
with a larger vertical scale. 
The observed sharp breakdown of agreement at $\sqrt{s_{NN}} = 7.7$ GeV implies
that one or more of the aforementioned assumptions no longer hold below 11.5 GeV.
\begin{figure}[!htb]
\begin{minipage}[c]{0.58\textwidth}
  \includegraphics[width=\textwidth]{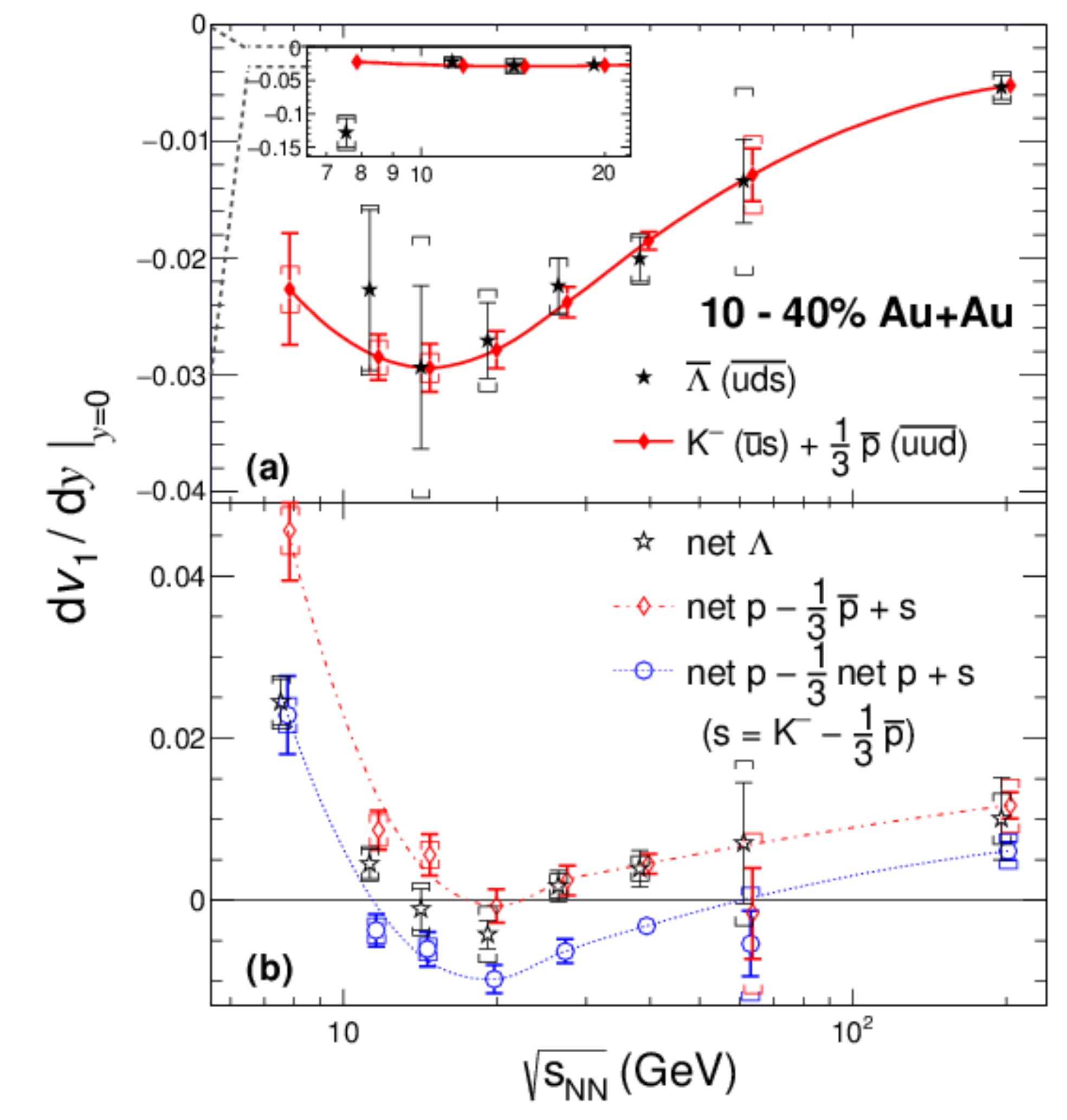}
  \label{fig1}
\end{minipage}
\begin{minipage}[c]{0.40\textwidth}
\caption{ (Color online) Directed flow slope ($dv_1/dy$) versus $\sqrt{s_{NN}}$ for intermediate centralities ($10-40\%$)~\cite{STAR2}. 
Panel (a) compares the observed $\Lambda$ slope with the prediction of the coalescence
sum rule for produced quarks. The inset shows the same comparison where the vertical scale is zoomed-out; this allows the observed flow for the lowest energy ($\sqrt{s_{NN}} = 7.7$ GeV) to be seen. Panel (b) presents two further sum-rule
tests, based on comparisons with net-$\Lambda$ measurements. The solid and dotted lines are smooth curves to guide the eye.}
\end{minipage}
\vspace{-0.5cm}
\end{figure}

In the limit of low $\sqrt{s_{NN}}$, most $u$ and $d$ quarks are presumably transported, whereas in the limit of high $\sqrt{s_{NN}}$, most of $u$ and $d$ are produced. 
In Fig.~1(b), we exploit net $\Lambda(uds)$ to test two coalescence sum rule scenarios which are expected to bracket the observed $dv_1/dy$ for a baryon containing transported quarks. The first compared data operation (red diamond
markers) consists of net proton ($uud$) minus $\bar u$ plus $s$, where $\bar u$ is estimated from $\frac{1}{3}{\bar p}$, and the $s$ quark flow is obtained from $K^-({\bar u}s) - \frac{1}{3}{\bar p}({\bar u}{\bar u}{\bar d})$. Here we assume that a produced $u$ quark in net $p$ is replaced with an $s$ quark. This sum-rule calculation agrees closely with the net-$\Lambda$ measurement at $\sqrt{s_{NN}} = 19.6$ GeV and above, remains moderately close
at $14.5$ and $11.5$ GeV, and deviates significantly only at $7.7$ GeV. With decreasing beam energy, the number of transported quarks per net proton increases (as shown in Fig.~4), and there is an increasing departure from the assumption that a produced $u$ quark is removed by keeping the term (net $p - \frac{1}{3}{\bar p}$).
The second coalescence operation in Fig.~1(b) corresponds to $\frac{2}{3}$ net proton plus $s$ (blue circle markers). In this case, we assume that the constituent quarks of net protons are dominated by transported quarks in the limit of low beam energy, and that one of the transported
quarks is replaced by $s$. This approximation only seems to hold at $\sqrt{s_{NN}} = 7.7$ GeV, and breaks down as the beam energy increases. 

\begin{figure}[!htb]
\begin{minipage}[c]{0.48\textwidth}
  \includegraphics[width=\textwidth]{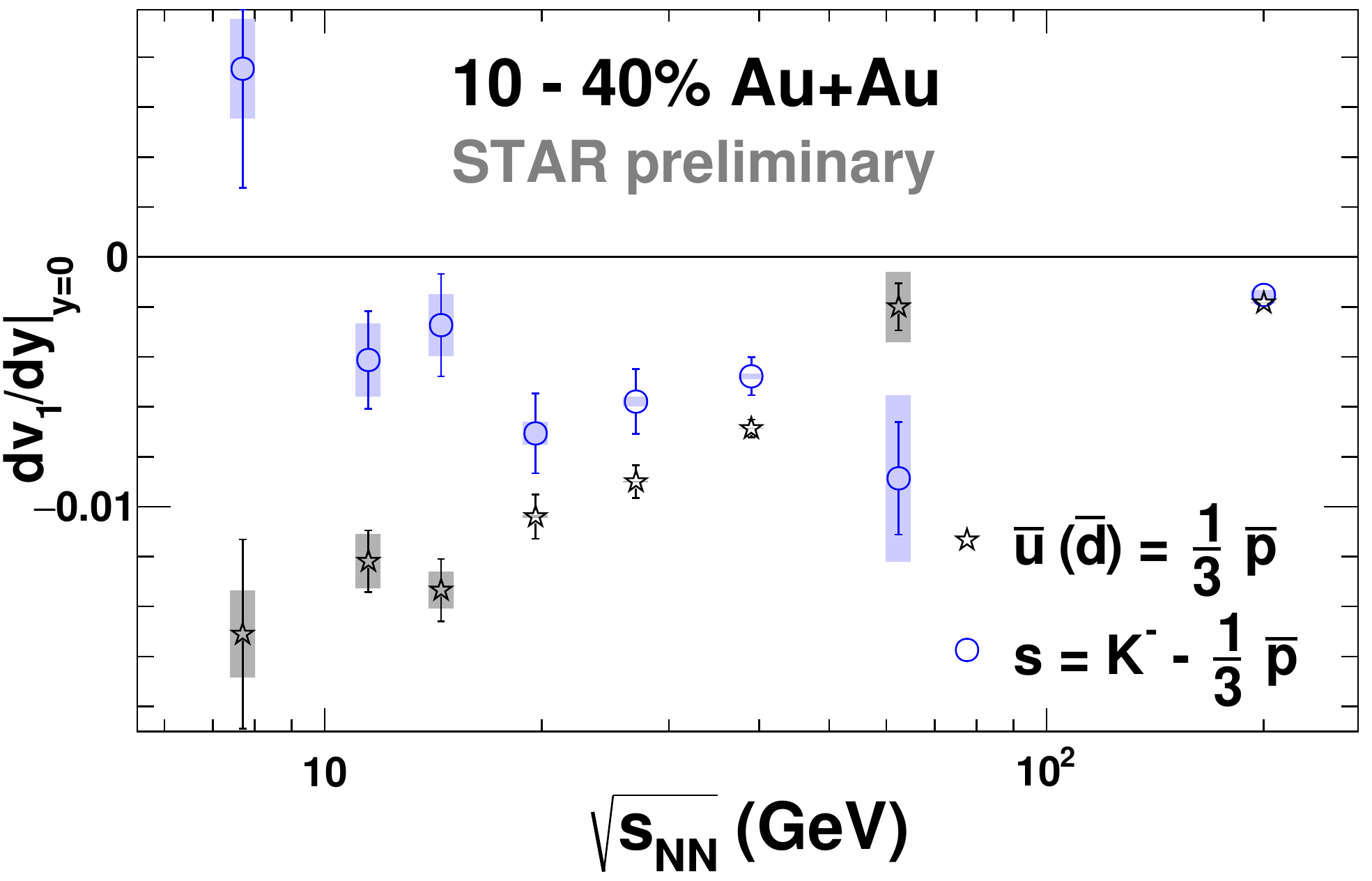}
  \caption{(Color online) Directed flow slope ($dv_1/dy$) of ${\bar u} ({\bar d})$ and $s$ quarks versus $\sqrt{s_{NN}}$ for intermediate centralities ($10-40\%$).}
  \label{fig2}
\end{minipage}
\begin{minipage}[c]{0.48\textwidth}
  \includegraphics[width=\textwidth]{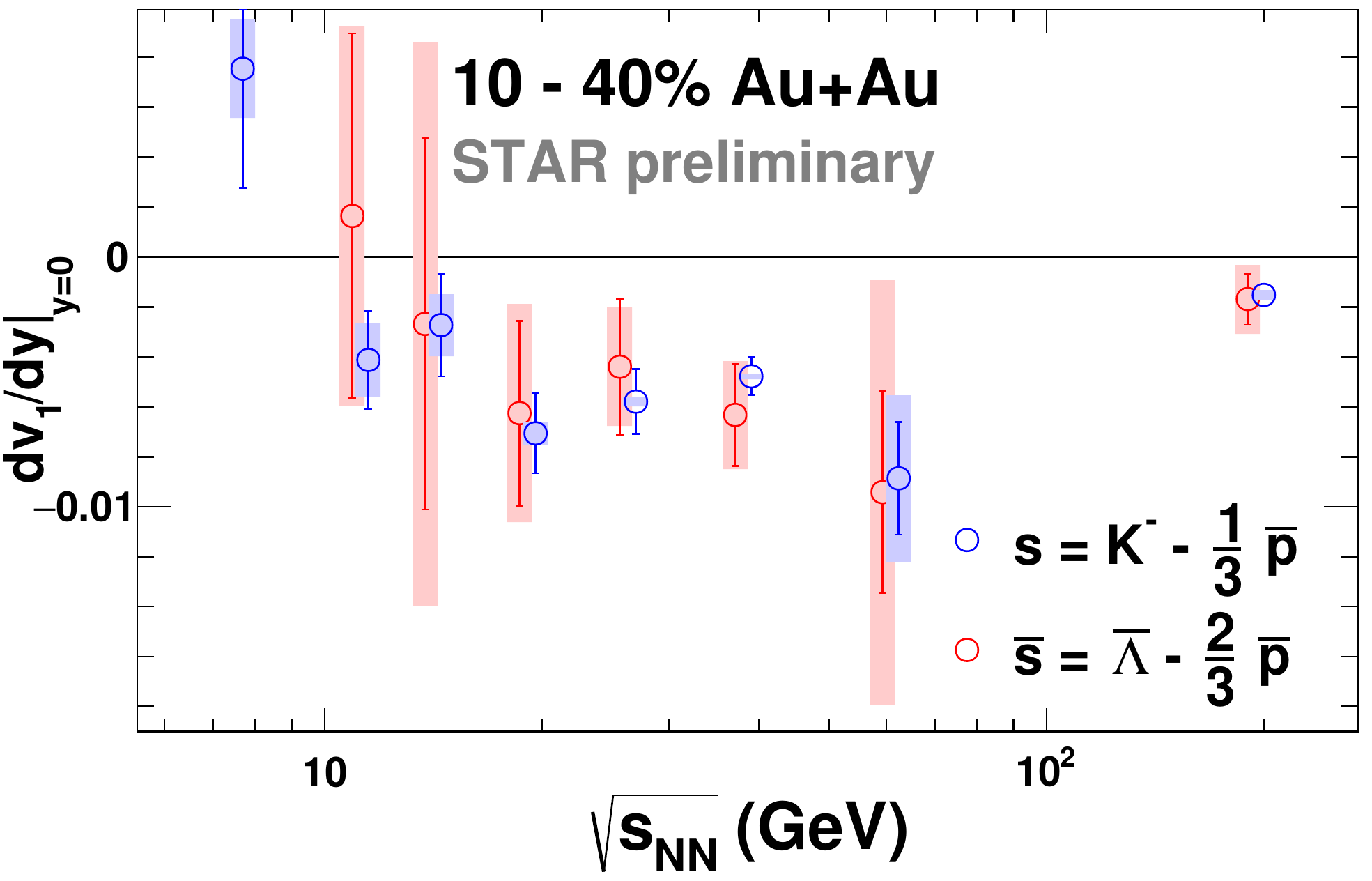}
  \caption{(Color online) Directed flow slope ($dv_1/dy$) of $s$ and ${\bar s}$ quarks versus $\sqrt{s_{NN}}$ for intermediate centralities ($10-40\%$).}
  \label{fig3}
\end{minipage}  
\end{figure}

With the assumption of the coalescence sum rule, we can further extract $dv_1/dy$ of constituent quarks. Figures~2 and 3 present directed flow slope ($dv_1/dy$) of produced quarks as function of $\sqrt{s_{NN}}$ for intermediate centralities ($10-40\%$). $dv_1/dy$ of the produced $u$($d$, $\bar u$, $\bar d$) quarks is approximated with 
$\frac{1}{3}{\bar p}({\bar u}{\bar u}{\bar d})$, whereas the $s$ quark flow
is obtained from $K^-({\bar u}s) - \frac{1}{3}{\bar p}$, and
$\bar s$ from $\bar \Lambda({\bar u}{\bar d}{\bar s}) - \frac{2}{3}{\bar p}$.
At $\sqrt{s_{NN}} = 200$ GeV, the produced $u$($d$, $\bar u$, $\bar d$) and $s$ quarks have very similar $v_1$ slopes, and they deviate towards lower beam energies. The consistency between $s$ and $\bar s$ holds within uncertainties for almost all studied beam energies, except at $\sqrt{s_{NN}} = 7.7$ GeV, where $dv_1/dy$ of $\bar s$ is $-0.097\pm 0.023({\rm stat.})\pm 0.026({\rm syst.})$, far off the scale. 

\begin{figure}[!htb]
\begin{minipage}[c]{0.48\textwidth}
  \includegraphics[width=\textwidth]{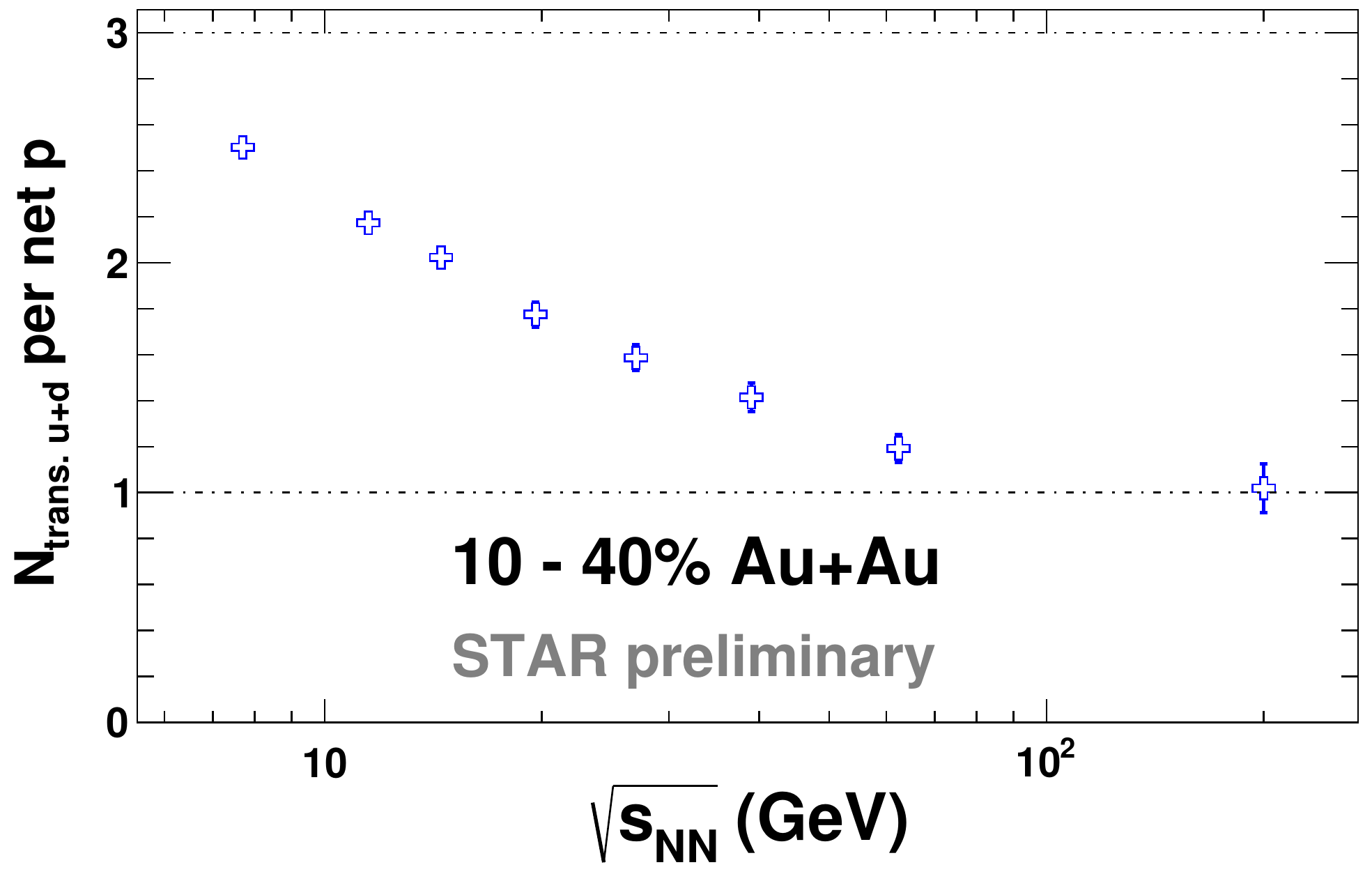}
  \caption{(Color online) Number of transported $u$ and $d$ quarks per net proton versus $\sqrt{s_{NN}}$ for intermediate centralities ($10-40\%$).}
  \label{fig5}
\end{minipage}
\begin{minipage}[c]{0.48\textwidth}
  \includegraphics[width=\textwidth]{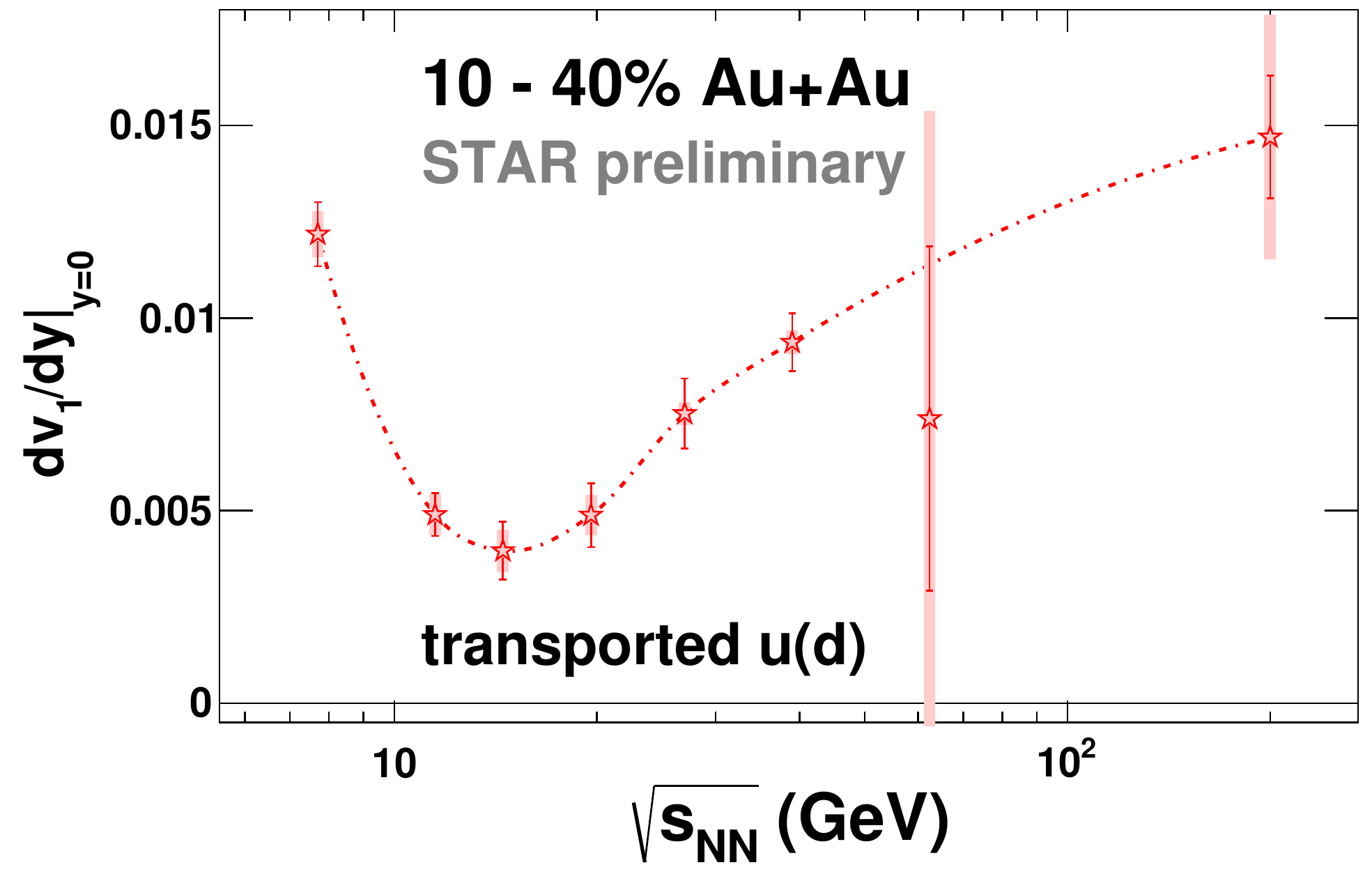}
    \caption{(Color online) Directed flow slope ($dv_1/dy$) of transported $u(d)$ quarks versus $\sqrt{s_{NN}}$ for intermediate centralities ($10-40\%$).}
  \label{fig6}
\end{minipage}  
\end{figure}

Next, we estimate the number of transported quarks per net proton. Assuming that the chemical potential of $u(d)$ quarks is one third of that of baryons, $\mu_{u(d)} = \mu_{\rm B}/3$, 
we derive from thermal equilibrium that the number of transport quarks per net proton is $N_{\rm {trans.}\,u+d} = 3\times(1-e^{-2\mu_{u(d)}/T_{\rm ch}})/(1-r)$~\cite{JFL}. Here $T_{\rm ch}$ is the chemical freeze-out temperature, and $r\propto e^{-6\mu_{u(d)}/T_{\rm ch}}$ is the aforementioned ratio of observed $\bar p$ to $p$ yield.
In the limit of low $\sqrt{s_{NN}}$ or high $\mu_{\rm B}$, $N_{\rm {trans.}\,u+d}$ is close to three, whereas in the limit of high 
$\sqrt{s_{NN}}$ or low $\mu_{\rm B}$, $N_{\rm {trans.}\,u+d}$ approaches unity. These features are confirmed by Fig.~4 that shows the number of transported quarks per net proton versus $\sqrt{s_{NN}}$ for intermediate centralities ($10-40\%$). The values of $\mu_{\rm B}$ and $T_{\rm ch}$ are based on previous STAR measurements~\cite{muB}. Quoted errors are statistical uncertainties only.

Finally, the $v_1$ slope of transported quarks is obtained by removing produced quarks from net protons: $v_{1\,{\rm trans.}\,u(d)} = [v_{1\,{\rm net}\,p}-(3-N_{\rm {trans.}\,u+d})\times v_{1\,{\bar u}({\bar d})}]/N_{\rm {trans.}\,u+d}$. Figure~5 illustrates $dv_1/dy$ of transported $u(d)$ quarks versus $\sqrt{s_{NN}}$ for intermediate centralities ($10-40\%$), which is positive for all the beam energies under study, and demonstrates a minimum at $\sqrt{s_{NN}} \approx 14.5$ GeV. Compared with previous STAR results on net-proton $v_1$, these data on the constituent quark level provide a stronger evidence of the softening of equation of state, though a definite interpretation still requires further theoretical inputs.

\section{Summary}
STAR has published directed flow results for ten particle species in Au+Au collisions at eight beam energies. These data enable us to test the coalescence sum rule on the constituent quark level, for both produced and transported quarks.
The observed pattern of scaling behavior
for produced quarks at and above $\sqrt{s_{NN}} =11.5$ GeV, with
a breakdown at 7.7 GeV, warrants further study.
Two coalescence sum rule scenarios have been explored to bracket the observed $dv_1/dy$ of net $\Lambda$ that contains transported quarks.
Assuming the validity of the coalescence picture, we have extracted the $v_1$ slopes of produced $u$($d$, $\bar u$ and $\bar d$), $s$ and $\bar s$ quarks, as well as transported $u$($d$) quarks as function of $\sqrt{s_{NN}}$
for intermediate centralities.
Among produced quarks, $u$ and $s$ have similar $dv_1/dy$ at 200 GeV, and deviate towards lower beam energies. The $v_1$ slopes of $s$ and $\bar s$ are consistent with each other, except at 7.7 GeV. $dv_1/dy$ of transported quarks is positive for all the beam energies under study, and supports the softening of equation of state with a minimum at $\sqrt{s_{NN}} \approx 14.5$ GeV.
These energy-dependent measurements will be enhanced after STAR
acquires greatly increased statistics using upgraded detectors
in Phase II of the RHIC Beam Energy Scan~\cite{BES}.

\label{summary}

\section*{Acknowledgments}
This work is supported by by the US Department
of Energy under Grant No. DE-FG02-88ER40424.

%% References


\bibliographystyle{elsarticle-num}
\bibliography{<your-bib-database>}

\end{document}